\documentclass[a4paper]{PoS}
\pdfoutput=1
\usepackage{amsmath}
\usepackage[utf8]{inputenc}
\title{Phenomenology with Lattice NRQCD b Quarks.}

\ShortTitle{Phenomenology with Lattice NRQCD b Quarks.}
\author{\speaker{Brian Colquhoun}$^a$, Christine T. H. Davies$^a$, Rachel J. Dowdall$^b$, Jonna Koponen$^a$, G. Peter Lepage$^c$ and Andrew T. Lytle$^a$\\
        \llap{$^a$}SUPA, School of Physics \& Astronomy, University of Glasgow, Glasgow, G12 8QQ, UK\\
        \llap{$^b$}DAMTP, University of Cambridge, Wilberforce Road, Cambridge, CB3 0WA, UK\\
        \llap{$^c$}Laboratory of Elementary-Particle Physics, Cornell University, Ithaca, NY 14853, USA\\
        E-mail: \email{brian.colquhoun@glasgow.ac.uk}}
\author{HPQCD Collaboration\thanks{URL: http://www.physics.gla.ac.uk/HPQCD}}

\abstract{The HPQCD collaboration has used radiatively-improved NonRelativistic QCD (NRQCD) for $b$ quarks in bottomonium to determine the decay rate of $\Upsilon$ and $\Upsilon^\prime$ mesons to leptons in lattice QCD. Using time-moments of vector bottomonium current-current correlators, we are also able to determine the $b$ quark mass in the $\overline{\mathrm{MS}}$ scheme. We use the same NRQCD $b$ quarks and Highly Improved Staggered Quark (HISQ) light quarks -- with masses down to their physical values -- to give a complete picture of heavy-light meson decay constants including those for vector mesons. We also study the semileptonic $B\rightarrow\pi\ell\nu$ decay at zero recoil to show that lattice QCD is consistent with the soft pion theorem for this decay: $f_0(q^2_{\mathrm{max}})=f_B/f_\pi$ in the massless pion limit. Finally, we present preliminary results for the $B_c \rightarrow \eta_c \ell \nu$ semileptonic decay form factors. This is a showcase for the comparison of results for NRQCD $b$ quarks with those from HISQ $b$ quarks (both with HISQ $c$ quarks). We give the first 3-point results from our `heavy HISQ' programme, which will allow us to improve the normalisation of NRQCD-HISQ currents for other calculations.}

\FullConference{The 33rd International Symposium on Lattice Field Theory\\
		 14 -18 July  2015\\
		 Kobe International Conference Center, Kobe, Japan}

\begin{document}

\section{Introduction}
Lattice NRQCD is a computationally inexpensive formalism that allows us to perform calculations for $b$ quarks at their physical mass on relatively coarse lattices. Radiative improvements to the action reduce systematic uncertainties~\cite{Dowdall:2011wh,Hammant:2013sca}. In contrast, the HISQ action~\cite{Follana:2006rc} requires multiple values of quark masses larger than that of the $c$ mass and an extrapolation to $m_b$. The comparison of the two provides important tests of lattice QCD systematic errors, and we will discuss several such comparisons here. We use the HISQ 2+1+1 gluon field ensembles provided by the MILC collaboration at a variety of lattice spacing values that include degenerate light ($u/d$) sea quarks with masses down to their physical value~\cite{Bazavov:2012xda}.

\section{Bottomonium}
The $\Upsilon$ or $\Upsilon^{\prime}$ leptonic width can be expressed in terms of the square of its decay constant, $f_{\Upsilon^{(\prime)}}$, by,
\begin{equation}
\Gamma\left(\Upsilon^{(\prime)}\rightarrow e^+e^-\right)=\frac{4\pi}{3}\alpha^2_{\mathrm{QED}}e^2_b\frac{f^2_{\Upsilon^{(\prime)}}}{M_{\Upsilon^{(\prime)}}}, \qquad f_{\Upsilon^{(\prime)}}M_{\Upsilon^{(\prime)}}\delta_{ij} = \langle 0 |J_{V,i}|\Upsilon^{(\prime)}_j \rangle,
\end{equation}
where $\alpha_{\mathrm{QED}}$ is the fine structure constant, $e_b$ is the $b$ quark charge and $M_{\Upsilon^{(\prime)}}$ is the mass of the $\Upsilon^{(\prime)}$.

To represent the vector $b\bar{b}$ current, $J_{V,i}$, on the lattice, we take leading and subleading operators,
\begin{equation}
J^{(0)}_{V,\mathrm{NRQCD},i}=\chi^\dagger\sigma_i\psi, \qquad J^{(1)}_{V,\mathrm{NRQCD},i}=\chi^\dagger\sigma_i\frac{\hat{\Delta}^{(2)}}{(am_b)^2}\psi, \qquad J_{V,i}=Z_V(J^{(0)}_{V,\mathrm{NRQCD},i}+k_1J^{(1)}_{V,\mathrm{NRQCD},i}),
\end{equation}
for improved NRQCD $b$ quarks through NLO in the $b$ quark velocity. We determine $k_1$ and  $Z_V$ nonperturbatively using time-moments of the correlator and high-order continuum QCD perturbation theory~\cite{Colquhoun:2014ica}.

\begin{figure}
\centering{
\includegraphics[width=0.6\linewidth]{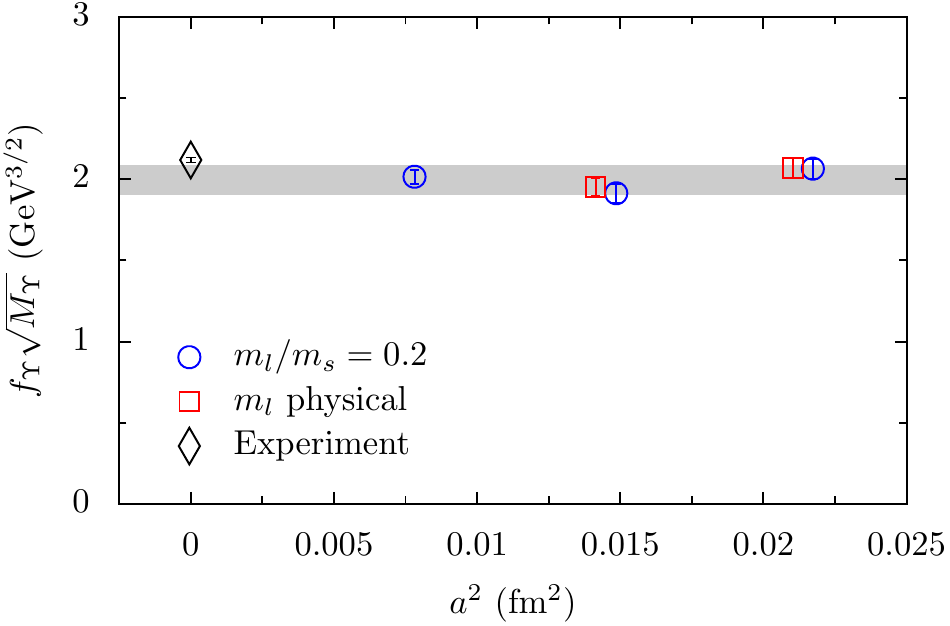}
\caption{Results for the hadronic parameter $f_{\Upsilon}\sqrt{M_\Upsilon}$~\cite{Colquhoun:2014ica}. The blue points are for ensembles with $m_l/m_s=0.2$ and the red points are for ensembles with physical light quarks. The grey band shows our physical value with all sources of uncertainty included. The result using the experimental value of the leptonic width and mass is shown as the black diamond, corresponding to $f_\Upsilon\sqrt{M_{\Upsilon}}=2.119(14)~\mathrm{GeV}^{3/2}$.}
\label{fig:upsilon-decay}
}
\end{figure}

Figure~\ref{fig:upsilon-decay} shows our results for the hadronic parameter $f_\Upsilon\sqrt{M_{\Upsilon}}$ on very coarse $(a\approx0.15~\mathrm{fm})$, coarse $(a\approx0.12~\mathrm{fm})$ and fine $(a\approx0.09~\mathrm{fm})$ lattices at multiple sea light quark masses. Using the value of $M_\Upsilon$ from experiment, we find $f_\Upsilon=0.649(31)~\mathrm{GeV}$ from our fit to the form discussed in~\cite{Colquhoun:2014ica}. We can also use the experimental value of the $\Upsilon$ mass along with $\alpha_{\mathrm{QED}}$ to get the leptonic width and we obtain $\Gamma(\Upsilon \rightarrow e^+e^-) = 1.19(11)~\mathrm{keV}$. This result is within $1.5\sigma$ of the experimental value of $\Gamma(\Upsilon \rightarrow e^+e^-) = 1.340(18)~\mathrm{keV}$ , corresponding to $f_{\Upsilon}=0.689(5)~\mathrm{GeV}$.

To calculate the decay constant for the $\Upsilon^{\prime}$ we take ratios of amplitudes to cancel uncertainties arising from the matching factor $Z_V$. Using our result for $f_\Upsilon$, we obtain $f_{\Upsilon^{\prime}}=0.481(39)~\mathrm{GeV}$ and $\Gamma(\Upsilon^\prime \rightarrow e^+e^-) = 0.69(9)~\mathrm{keV}$. This is again in good agreement with the experimental value of $\Gamma(\Upsilon^\prime \rightarrow e^+e^-) = 0.612(11)~\mathrm{keV}$.

\section{$b$ Quark Mass}
\begin{figure}
\centering{
\includegraphics[width=0.6\linewidth]{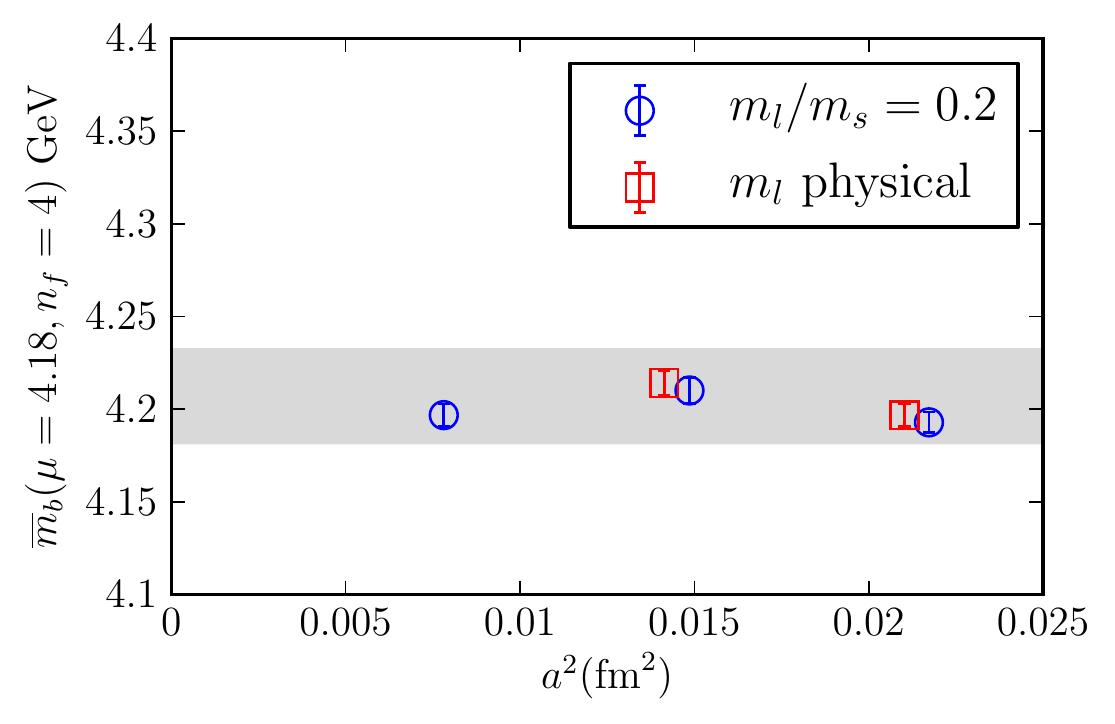}
\caption{Results for $\overline{m}_b$ against the square of the lattice spacing~\cite{Colquhoun:2014ica}. The blue points are for ensembles with $m_l/m_s=0.2$ and the red points are for ensembles with physical light quarks. The grey band shows the result of our fit to these points.}
\label{fig:mb}
}
\end{figure}

The relationship between correlator time-moments and continuum QCD perturbation theory also allows us to determine the $b$ quark mass from our results. In figure~\ref{fig:mb} we show our results for $\overline{m}_b$ extracted from moment $n=18$ (in our notation where we calculate even moments $4,6,8,10,\ldots$), with the grey band showing our final value of $4.207(21)~\mathrm{GeV}$. Converting to 5 flavors and including systematic uncertainties as detailed in~\cite{Colquhoun:2014ica} we find $\overline{m}_b(\overline{m}_b, n_f=5)=4.196(23)~\mathrm{GeV}$. This agrees well with the result of our `heavy HISQ' analysis in which we study pseudoscalar correlators made of HISQ quarks with a variety of masses larger than $m_c$ and extrapolate to the mass of the $b$ to find $\overline{m}_b(\overline{m}_b, n_f=5)=4.164(23)~\mathrm{GeV}$~\cite{Chakraborty:2014aca,McNeile:2010ji}.

\section{$B$ and $B^*$ Mesons}
We have recently extended our determination of the $B$ and $B_s$ decay constants using improved NRQCD $b$ quarks and HISQ light quarks~\cite{dowdall-bmeson}. The value of the ratio of vector to pseudoscalar decay constants -- and whether it is larger or smaller than 1 -- has been a controversial topic for some years. 

To reduce systematic errors we determine a ratio,
\begin{equation}
\frac{f_{B^*_q}\sqrt{M_{B^*_{q}}}}{f_{B_q}\sqrt{M_{B_{q}}}} = \left(1+z_0\alpha_s\right)\frac{\left(\Phi^{(0)}_{V_i}+\Phi^{(1)}_{V_i}\right)}{\left(\Phi^{(0)}_{A_0}+\Phi^{(1)}_{A_0}\right)},
\end{equation}
from which we can readily extract $f_{B^*_q}/f_{B_q}$. $\Phi^{(0)}$ and $\Phi^{(1)}$ represent the leading and subleading currents in an expansion in powers of $\Lambda/m_b$. Details of this calculation are given in~\cite{Colquhoun:2015oha}.

We find $f_{B^*_s}/f_{B_s}=0.953(23)$ and $f_{B^*}/f_{B}=0.941(26)$, both of which are $2\sigma$ \textit{below} 1. For $f_{B^*_c}/f_{B_c}$ we obtain $0.988(27)$. We are also able to compare $f_{B_c}$ using improved NRQCD $b$ quarks to that using heavy HISQ $b$ quarks~\cite{McNeile:2012qf}, where we have an absolute normalisation of the current. Good agreement is found between these methods, similar to that between NRQCD-HISQ and HISQ-HISQ for $f_{B_s}$~\cite{dowdall-bmeson,McNeile:2011ng}.

\begin{figure}
\centering{
\includegraphics[width=0.69\textwidth]{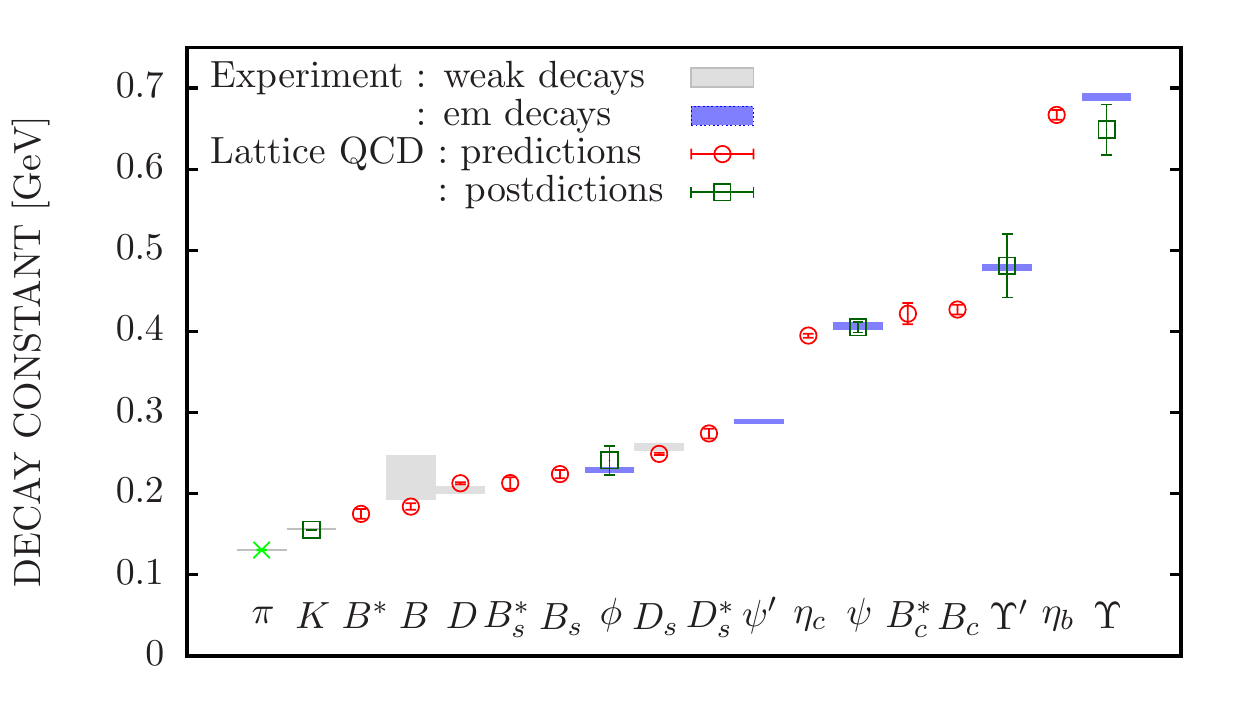}
\caption{Summary of lattice QCD results of well-characterised mesons~\cite{Colquhoun:2015oha}. The grey and blue bands show values extracted from experimental annihilation rates for pseudoscalars (to $W$s) and vectors (to photons) respectively. }
\label{fig:decay-summary}
}
\end{figure}

\begin{figure}
\centering{
\includegraphics[width=0.63\textwidth]{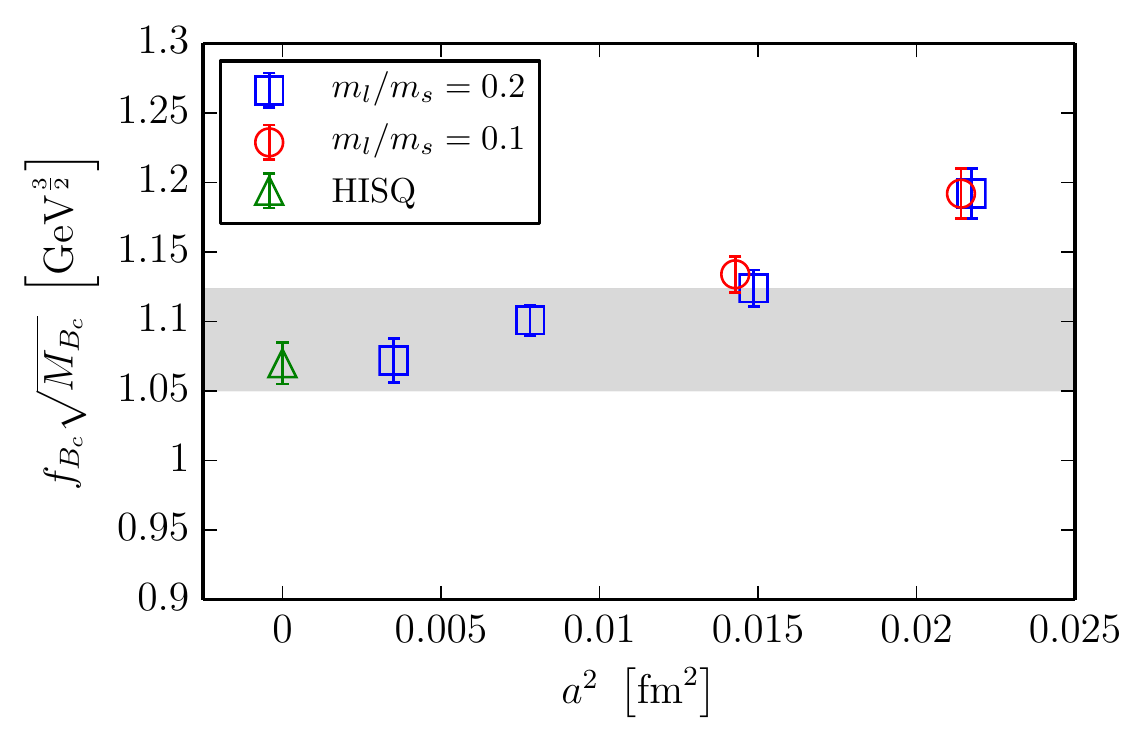}
\caption{Results for $f_{B_c}$ using improved NRQCD $b$ quarks (blue squares and red circles) in comparison with the physical result for HISQ $b$ and $c$ quarks (green triangle). The finest lattice on which we used the NRQCD formalism is our new result. It does not contribute to the fit result shown by the grey band, which comes from the coarse three lattice spacing values; see~\cite{Colquhoun:2015oha}.}
\label{fig:fBc}
}
\end{figure}

Figure~\ref{fig:decay-summary} shows a summary plot of decay constants determined from lattice QCD and ordered by magnitude. This includes all of the decay constant determinations discussed here. An updated plot from our $f_{B_c}$ calculation now including a preliminary NRQCD result at $a\approx0.06~\mathrm{fm}$ is shown in figure~\ref{fig:fBc}. For comparison, the green triangle is the physical result using the HISQ formalism for both the $b$ and $c$~\cite{McNeile:2012qf}.

\section{Semileptonic Decays}
\subsection{$B\rightarrow\pi\ell\nu$ at Zero Recoil}
The Cabibbo-Kobayashi-Maskawa matrix element $V_{ub}$ can be determined through the semileptonic process $B \rightarrow \pi \ell \nu$ by comparing theoretically determined form factors and experimental rates. We give the first lattice QCD results including $u/d$ quarks going down to their physical values for the scalar form factor $f_0$ at zero recoil. A full discussion of this calculation can be found in ref.~\cite{Colquhoun:2015mfa}.

\begin{figure}[t]
\centering{
\includegraphics[width=0.6\linewidth]{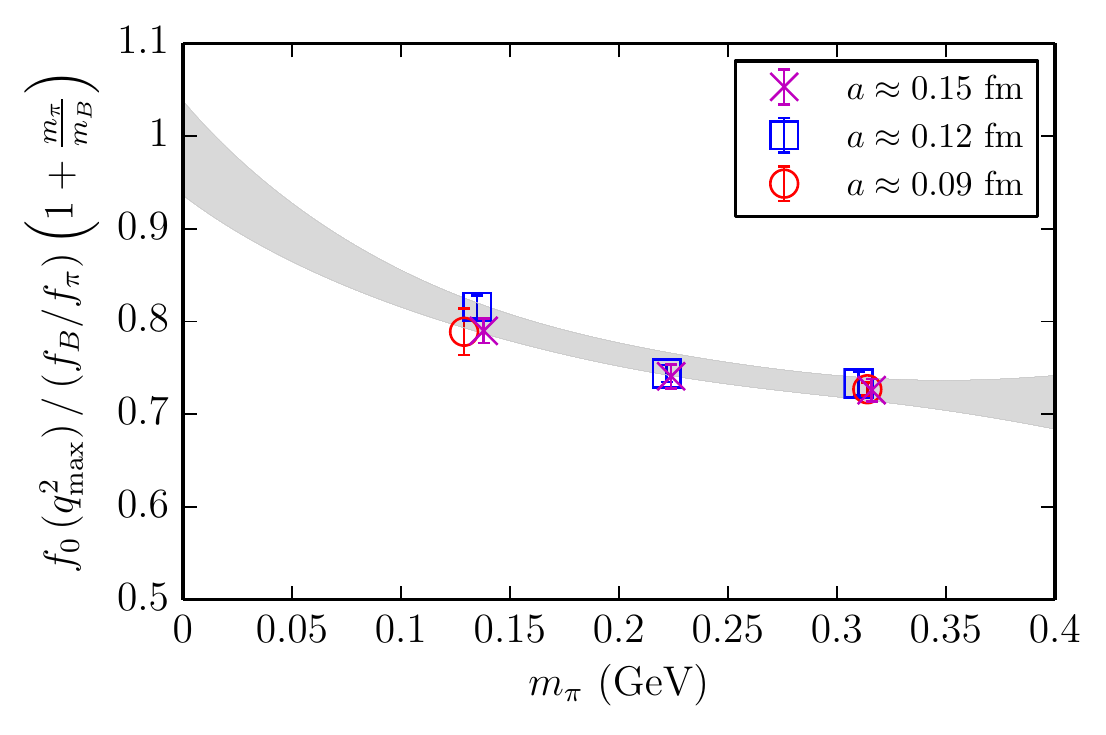}
\caption{Results for of the ratio of $f_0(q^2_{\mathrm{max}}) \times (1+m_\pi/m_B)$ to $f_B/f_\pi$~\cite{Colquhoun:2015mfa}. The grey band shows our final fit result, in agreement with 1 at $m_\pi=0$.}
\label{fig:soft-pion}
}
\end{figure}

We show our results for $f_0\left(q^2_{\mathrm{max}}\right)\left(1+m_\pi/m_B\right)\times\left[f_\pi/f_B\right]$ against $m_\pi$ using improved NRQCD $b$ quarks and HISQ light quarks in figure~\ref{fig:soft-pion}, with points spanning very coarse to fine lattices. The grey band shows our fit result as a function of $a$ and $m_\pi$ from which we obtain,
\begin{equation}
\left.\frac{f_0(q^2_{\mathrm{max}})}{f_B/f_\pi}\right\rvert_{m_\pi=0}=0.987(51),
\end{equation}
in agreement with the soft pion theorem result of 1. We show in~\cite{Colquhoun:2015mfa} that staggered quark chiral perturbation theory behaves very like continuum chiral perturbation theory for this range of $m_\pi$ values. At the experimental value $m_\pi=135~\mathrm{MeV}$ we find,
\begin{equation}
\left.f_0\left(q^2_{\mathrm{max}}\right)\right\rvert_{B^+ \rightarrow \pi^0}=1.108(22)(24),
\end{equation}
where the uncertainties are from statistics/fitting and from our value of $f_{B^+}$, which includes uncertainties from current matching. This can be compared to results from other lattice QCD calculations that extrapolate down in $m_\pi$ from unphysical values~\cite{Lattice:2015tia,Flynn:2015mha}.

\subsection{$B_c\rightarrow\eta_c\ell\nu$}
The semileptonic $B_c \rightarrow \eta_c \ell \nu$ decay is a physical process that does not involve light quarks. This enables us to provide an accurate comparison of results using improved NRQCD $b$ quarks and HISQ $c$ quarks with those of heavy HISQ quarks of varying mass combined with HISQ $c$ quarks. This is akin to the comparisons we have done for $m_b$, $f_{B_s}$ and $f_{B_c}$ discussed above.

The scalar form factor has an absolutely normalised current for HISQ quarks, whereas for the NRQCD case it is constructed from current contributions that must be combined and renormalised. A test across a large range of $q^2$ values -- in which the different contributions vary in size -- can then provide a strong test of the normalisation condition for the NRQCD-HISQ currents for other calculations (such as $B \rightarrow D^{(*)}\ell\nu$). For $B_c \rightarrow \eta_c \ell \nu$ we can cover the full range of $q^2$ values on superfine $(a\approx0.06~\mathrm{fm})$ and ultrafine $(a\approx0.045~\mathrm{fm})$ for HISQ quarks without large discretisation errors. We use the improved NRQCD formalism on fine and superfine lattices.

\begin{figure}
\centering{
\includegraphics[width=0.7\linewidth]{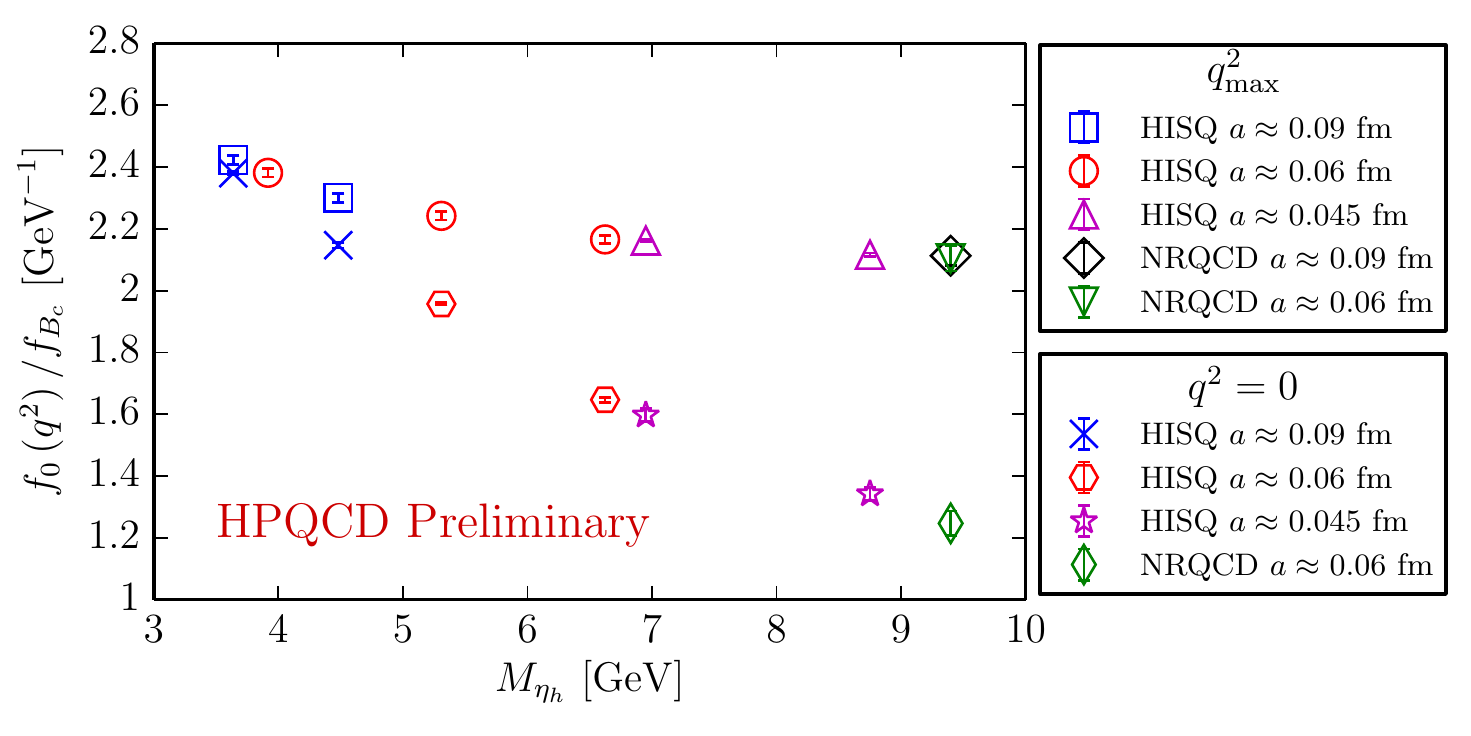}
\caption{Results for $f_0(q^2)/f_{B_c}$ for $q^2=q^2_{\mathrm{max}}$ (upper points) and for $q^2=0$ (lower points) against the pseudoscalar heavionium mass, $M_{\eta_h}$. The results for the rightmost value of $M_{\eta_h}$ use NRQCD $b$ quarks and HISQ $c$ quarks on the fine and superfine lattices. All other points show results on the fine, superfine and ultrafine ensembles using the HISQ formalism for both valence quarks.}
\label{fig:NRQCD_HISQ_Bc_etac}
}
\end{figure}

Figure~\ref{fig:NRQCD_HISQ_Bc_etac} compares results from NRQCD $b$ and HISQ $c$ quarks to those from HISQ heavy quarks $h$ and HISQ $c$ quarks as a function -- in the HISQ-HISQ case -- of the pseudoscalar heavy-heavy mass. Good agreement is seen between NRQCD-HISQ, which is perturbatively normalised, and HISQ-HISQ results for both $q^2=q^2_{\mathrm{max}}$ and $q^2=0$ despite very different dependence on $M_{\eta_h}$. No discretisation errors are visible when $f_0$ is normalised by $f_{B_c}$. We expect $f_0$ at fixed $\eta_c$ energy to scale with $M_{\eta_h}$ in a similar way to $f_{B_c}$. The ratio $f_{0}(q^2_{\mathrm{max}})/f_{B_c}$ would then tend to a constant as figure~\ref{fig:NRQCD_HISQ_Bc_etac} shows.

\section{Conclusions}
We have described several recent accurate results for the properties of mesons containing $b$ quarks. These have enabled us to fill out the picture of meson decay constants from lattice QCD. For $m_b$, $f_{B_s}$ and $f_{B_c}$ we have compared results using NRQCD $b$ quarks with those from heavy HISQ quarks extrapolating to $b$. The two methods have completely different systematic uncertainties and so their agreement is non-trivial confirmation of our understanding of those uncertainties. In ongoing work, we have shown here the first results for 3-point functions comparing heavy HISQ quarks to NRQCD for form factors for the $B_c \rightarrow \eta_c \ell \nu$ decay at both $q^2_{\mathrm{max}}$ and $q^2=0$. The good agreement seen is encouraging for our analysis, aiming to improve the normalisation of NRQCD-HISQ currents for a range of calculations.

\acknowledgments{This work was performed on the Darwin supercomputer as part of STFC's DiRAC facility.}

\bibliographystyle{modJHEP}
\bibliography{sources}

\end{document}